\documentclass{article}
\usepackage{graphicx} 

\title{The Combination Problem for Relational Quantum Mechanics }
 \author{Emily Adlam  \thanks{Philosophy Department and Institute for Quantum Studies, Chapman University, Orange, CA92866, USA \texttt{eadlam90@gmail.com} }}
\date{\today}

     \usepackage[top=2.5cm, bottom=2.5cm, left=2.5cm, right=2.5cm]{geometry}
 
     \usepackage{setspace}
\begin{document}

\maketitle

Panpsychism is a theory of phenomenal consciousness which suggests that every physical system possesses some form of consiousness. One particularly important issue for panpsychism is the `combination problem,' referring to the fact that panpsychism must offer some account  of how the consciousnesses associated with individual fundamental particles combine to give rise to higher-level consciousnesses associated with beings like us. 

Relational quantum mechanics (RQM) is a proposed solution to the quantum measurement problem, originated by Carlo Rovelli. A central feature of RQM is that it attributes `relative facts' to every physical system, and therefore RQM faces a  problem somewhat similar to the panpsychist combination problem: it must offer some account of how the relative facts associated with individual fundamental particles are related to higher-level relative facts associated with beings like us. In this article, I use the existing literature on the panpsychist combination problem  as a starting point to think about how to address this structurally similar combination problem in RQM.  

I begin in section \ref{intro} with an introduction to relational quantum mechanics. Then in section \ref{comp} I discuss some similarities and differences between RQM's combination problem and panpsychism's combination problem. Finally section \ref{solutions} reviews a number of proposed solutions to panpsychism's combination problem, and assesses the prospects for a similar solution in the context of RQM. I will argue that overall the prospects for solving RQM's combination problem look better for RQM+CPL, the alternative version of RQM proposed in ref \cite{https://doi.org/10.48550/arxiv.2203.13342}, than for orthodox versions of RQM.

\section{Relational Quantum Mechanics \label{intro}}

Relational quantum mechanics\cite{1996cr,vanFraassen2010-VANRQM,sep-qm-relational,pittphilsci19664} is a proposed solution to the quantum measurement problem based on the idea that quantum states and the outcomes of quantum measurements are not absolute: they are relativized to individual observers. For example,  when Alice measures a system $B$, unitary quantum mechanics as applied  by an external observer tells us that  Alice and $B$ will form an entangled state $\sum_i c_i | O_i \rangle_B \otimes |C_i \rangle_A$, where $\{ |O_i \rangle_B \}$ are states of $B$ corresponding to the possible outcomes and each $|C_i \rangle$ represents the state of Alice corresponding to her witnessing the outcome associated with state $|O_i\rangle$. Relational quantum mechanics stipulates that this interaction creates a `relative fact' about the outcome $O_i$ relative to Alice, determined probabilistically according to the Born rule; so Alice will experience the measurement as having a single definite outcome, even though from an external perspective it looks like the interaction has produced a superposition of all possible outcomes.

There exist  many versions of  this kind of `perspectival' solution to the measurement problem, including  neo-Copenhagen interpretations\cite{brukner2015quantum,Janas2021-JANUQR} and possibly QBism\cite{fuchs2023qbism}. What sets RQM apart is its insistence that any physical system whatsoever can play the role of an `observer' to which facts can be relativized - by contrast, other perspectival approaches suggest that this role can be played only by macroscopic sentient beings like ourselves. Thus in RQM we must extend the description  above to more general interactions. So when two arbitrary quantum systems $X$ and $Y$ interact, unitary quantum mechanics as applied  by some external observer tells us that they will  form an entangled state, which I will assume to be pure\footnote{Rovelli suggests that when the state is not pure, this means the measurement is not completed\cite{1996cr}. The problem of understanding exactly when relative facts are created in RQM\cite{Mucino2022-MUCARQ} is not yet fully solved, but I will not try to address it here.}. This state can be written in many different bases, but we will always be able to find at least one basis in which it takes the Schmidt form\cite{Preskilllecture} $\sum_i c_i | U_i \rangle_X \otimes | V_i \rangle_Y$, such that the sets $\{ | U_i \rangle_X \}$,$\{ | V_i \rangle_Y \}$ are orthonormal, and `cross terms' $| U_i \rangle_X \otimes |V_j \rangle_Y$ for $i \neq j$ all have coefficients equal to zero. Relational quantum mechanics then stipulates that this interaction creates a relative fact for $X$ specifying that system $Y$ has some value $V_i$ of the variable $V$, and a relative fact for $Y$  specifying that system $X$ has some value $U_i$ of the variable $U$, with both facts again selected probabilistically according to the Born rule.

Note that as emphasized by di Biagio and Rovelli\cite{pittphilsci19664} the relative facts produced in the interaction between $X$ and $Y$ are an intrinsic property of their interaction,  not of the state which an external observer would assign to them - indeed clearly this must be the case, since RQM allows that different external observers may assign different states to $X$ and $Y$. So it would be more correct to determine the basis of the relative facts established in an interaction from the interaction Hamiltonian, rather than from the resulting state defined relative to an external observer. However, there does not seem to be a well-established general procedure for inferring a basis from an interaction Hamiltonian, so in this paper I employ the more straightforwardly interpretable external state representation, on the understanding that the state assigned by an idealized external observer who is assumed to know what kind of  interaction has taken place is  a convenient way of describing the kind of relative facts that might emerge from this interaction\footnote{The use of the Schmidt decomposition to identify possible relative facts recalls its use in modal interpretations, in which it is often suggested that the Schmidt decomposition should determine possible value states\cite{sep-qm-modal}. Indeed, RQM could be regarded as belonging to the class of modal interpretations, and there exists a perspectival modal interpretation which bears similarities with RQM\cite{bene2001perspectival}. }. 

Evidently there are similarities between panpsychism and RQM, and   indeed,   RQM  could be regarded as a form    of panpsychism if  we assume the relative facts ascribed to  individual fundamental particles are associated with subjective experience. Rovelli himself suggests that `\emph{relationalism can be seen as a very mild form of panpsychism.}'\cite{Rovelli2021-ROVRAP-2}. However, a fully panpsychist version of RQM would make the problem I am discussing in this article harder, as it would impose   the constraint that  relative facts at every scale must  have features compatible with  our expectations about subjective experience. Indeed, Rovelli himself emphasizes that the parallel should not be taken too far, noting  that  `\emph{This is perhaps not panpsychism, because there is nothing specifically psychic or mental in the relational properties of a system with respect to another system}'\cite{Rovelli2021-ROVRAP-2}. Therefore  in this article I will focus on a version of RQM in which relative facts are not necessarily linked with subjective experience.  In this version of RQM, the term `observer' is used in quite a thin way: physical systems are called observers in virtue of having a set of relative facts associated with them, but they need not have consciousness, knowledge, or anything of that kind. In this article I will always use the term `observer' in this way;  I will use the term `sentient being' when I intend to refer to observers which are  conscious and have agency and knowledge.

 \subsection{Orthodox RQM and RQM+CPL}

In this article I will discuss two different versions of RQM. First, `orthodox RQM,' named thus both because it is the original version, and because it belongs to the class of `orthodox interpretations' of quantum mechanics\cite{cuffaro2021open}. Orthodox RQM insists that  \emph{all} physical facts must be relativized to an observer, so it is impossible for information to be exchanged between two different observers, since this would imply the existence of a non-relative fact about the connection between the two observers. Of course it may appear from the point of view of a third observer that information passes between two observers, but that is true only relative to the third observer, and does not imply any intrinsic connection between the relative facts of the first two observers. Thus in orthodox RQM there cannot be any systematic connections between the relative facts associated with different observers, and indeed some presentations of orthodox RQM suggest that we cannot even sensibly ask about such connections: `\emph{It is meaningless to compare events relative to different systems, unless this is done relative to a (possibly third) system}'\cite{dibiagio2021relational}.

Second `RQM+CPL,' as set out in ref \cite{https://doi.org/10.48550/arxiv.2203.13342}, in which RQM is supplemented with an additional postulate, Cross-Perspective Links (CPL): `\emph{In a scenario where some observer Alice measures a variable V of a system S, then provided that Alice does not undergo any interactions that destroy the information about V stored in Alice’s physical variables, if Bob subsequently measures the physical variable representing Alice’s information about the variable V, then Bob’s measurement result will match Alice’s measurement result.}' That is, in the scenario where  $X$ and $Y$   interact and form an entangled state  with Schmidt form  $\sum_i c_i | U_i \rangle_X \otimes | V_i \rangle_Y$, ordinary RQM tells us there will be a relative fact for $X$ specifying that system $Y$ has some value $V_i$ of the variable $V$;  so CPL stipulates that  if you now measure $X$ in the basis $\{ | U_i \rangle \}$, the result of your measurement will tell you this relative fact, i.e. if you obtain the result $U_i$ you will know that the value of $V$ for $Y$, relative to $X$, is $V_i$.  However, if you measure $X$ in some other basis, this interaction destroys the information about $Y$ stored in $X$'s physical variables, so the relative fact can no longer be learned from subsequent interactions.

  Evidently observers in RQM+CPL can share information, so it allows for systematic connections between relative facts associated with different observers. This  implies the existence of some non-relative  facts about the connections between different observers; for example, ref \cite{https://doi.org/10.48550/arxiv.2203.13342} suggests that RQM+CPL could be grounded on a set of non-relative physical facts about pointlike events which represent interactions between different physical systems. As argued by refs \cite{https://doi.org/10.48550/arxiv.2203.13342,https://doi.org/10.48550/arxiv.2203.16278}, this feature is essential because the  practice of science requires sharing information between observers, so a scientific theory which denies the possibility of such a thing is self-undermining. 
 
\section{Comparison \label{comp}}

In RQM there is a set of relative facts associated with `every physical system,' and this evidently includes   every fundamental particle, but it cannot  \emph{only} include fundamental particles, because  in order to reproduce  empirical results observed by human scientists RQM must attribute relative facts to human-sized systems as well\footnote{Note that I assume throughout this article that the most fundamental things are elementary particles. Quantum field theory indicates that this is not strictly true, but it is an appropriate assumption in this context, since the standard formulation of RQM relies crucially on the existence of well-defined, distinct physical systems like particles at the fundamental level. Generalizing RQM to a field theory context in which we do not have  distinct systems at the fundamental level comes with its own suite of problems, which I will not  address in this article.}. Thus, much like the panpsychist combination problem (PCP), we can identify RQM's combination problem (RCP): something must be said  about how the relative facts associated with individual fundamental particles are connected to the relative facts associated with macroscopic systems.

To get a sense of how deep the analogy between RCP and PCP goes, let us start with an obvious \emph{dis}analogy:  panpsychism and relational quantum mechanics are intended to solve  different problems, and naturally this will shape what we are looking for in solutions to their respective combination problems. 

Perhaps the most important goal of panpsychism is to solve the `hard problem of consciousness,' which refers to the fact that it is unclear how phenomenal consciousness could arise out of arrangements of microscopic non-conscious things\cite{Chalmers1995-CHAFUT}. The worry  is that if no reductive explanation of phenomenal consciousness can be given, we may be forced into a dualist view in which  macroscopic sentient beings must be regarded as a fundamental ontological category\cite{sep-dualism}, which many people find objectionable. Panpsychism seeks to avoid this outcome, and therefore solving PCP is essential to the goals of panpsychism -  for if a compelling solution to PCP cannot be found  it will be  equally unclear how macroscopic phenomenal consciousness could arise from a collection of microscopic \emph{conscious} things, so we may still be forced to treat  macroscopic sentient beings as a fundamental ontological category.
 
Meanwhile,  RCP may at first look like  a side issue which is not directly relevant to RQM's central goal of solving the measurement problem. However, one major reason we have a measurement problem in the first place is because we do not find it plausible that terms like `sentient being,' and `measurement'  should appear as unanalysable primitives in a fundamental theory of physics, so perspectival approaches which treat sentient beings as fundamental appear to be capitulating to the measurement problem rather than solving it. RQM aims to avoid this pitfall by treating all physical systems as observers: `\emph{the purpose of the multiplication of perspectives in RQM ... is that it offers a possible explanation to the key mystery of quantum physics: the apparent special role that (sentient beings) seem to have in the theory. RQM illuminates this mystery by denying that there is anything special in  (sentient beings)}'\cite{dibiagio2021relational}\footnote{In this quote Rovelli and di Biagio use the term `observers' refer to sentient beings, rather than in the thin sense in which the term is often used in RQM; so I have replaced it with `sentient beings' to avoid terminological confusion}.  This leads to what I will call the `central mandate' for RQM: in order for the strategy be successful, we must be able to fully formulate RQM's picture of reality without treating sentient beings, measurements, or anything of that kind as a fundamental ontological category. Hence  solving RCP is essential to the goals of RQM - for  if RCP cannot be solved in a compelling way, we will have to simply postulate macro-observers without any microphysical underpinnings, so we may still end up having to treat  macroscopic sentient beings as a fundamental ontological category.

 So  RQM and panpsychism share a central motivation, since both are partly driven by the desire to avoid anthropocentricism. Indeed, much of the quote above could have been written by a panpsychist: the purpose of the multiplication of consciousnesses in panpsychism is to explain the apparent special role that macroscopic sentient beings have in virtue of possessing consciousness, and panpsychism illuminates this mystery by denying that there is anything special in macroscopic sentient beings. In both cases  the combination problem is closely related to  this central motivation - it must be solved in order to show that the approach can be fully formulated without treating macroscopic sentient beings as  ontologically fundamental. Thus RCP in fact plays quite a similar role with respect to the measurement problem as PCP plays with respect to the hard problem.

\subsection{Empirical Enquiry}

There are also  similarities between RCP and PCP with regard to the extent to which they are  subject to empirical enquiry. Goff identifies a key difficulty for  PCP:  `\emph{It follows from the fact that we can introspect only one subject of experience that we cannot introspect how subjects of  experience qua subjects of experience are related, for to introspect how subjects of experience qua subjects of experience are related we would have to be able to introspect more than one subject of experience}'\cite{Goff2016-GOFTPB}. Of course,   can gain some indirect access to other people's experiences by asking questions and making inferences, so one might still think we could solve PCP empirically. However, we cannot  ask \emph{fundamental particles} questions about their subjective experience, so we can never access all the subjective experiences that panpsychism postulates within a given composite system in order to ascertain the relations between them.  

And much the same is true in   RQM: we have direct access only to our own relative facts, so we cannot directly observe the relations between relative facts associated with two different observers.  Indeed, in orthodox RQM we cannot access relative facts for systems other than ourselves at all. However,  RQM+CPL allows us  indirect access to   relative facts for other systems, so again, one might think that  in RQM+CPL we could  solve RCP empirically. But our ability to do this is limited, because CPL   stipulates that a relative fact for a given system becomes inaccessible if we measure the system in any basis other than the one recording this relative fact. Thus if we measure a fundamental particle we  change the values of the relative facts for any composite system to which it belongs, and meanwhile if we jointly measure the entire composite system we  change   the values of the relative facts for any of the fundamental particles making it up. So we can never check all the relative facts present in a given composite system  together in order to ascertain the relations between them; thus in both RCP and PCP, our ability to address the question empirically is limited.

\subsection{Versions of the Problem}
 
 To further explore the analogy between RCP and PCP, it is helpful to distinguish different facets of PCP. Chalmers identifies three\cite{Chalmers2016-CHATCP-6}:

\begin{itemize}

\item \textbf{Quality combination}: how do microqualities combine to form macroqualities?

\item \textbf{Structure combination}:  how does microexperiential and microphysical structure combine to yield macroexperiential structure?

\item \textbf{Subject combination}:  how do microsubjects combine to form macrosubjects?

\end{itemize}

 \subsubsection{Quality Combination}
 
 In RQM there is no  quality combination problem, because the relative facts employed in RQM need not be associated with anything like phenomenal consciousness or `qualities.'  There is however a somewhat analogous relative fact combination problem, since we need to know how relative facts for macro-observers are related  to relative facts for the micro-observers. But fortunately the relative fact combination problem seems  more straightforward than the quality combination problem. 
 
 Part of what makes the quality combination problem difficult  is the fact that phenomenal experience  seemingly cannot be defined in functional terms\cite{Chalmers1995-CHAFUT} - for example, the famous zombie argument  notes that it seems metaphysically possible that there could exist `zombies' which behave exactly the same way as human beings but which do not have phenomenal experience\cite{88fc19d1-07f0-3f02-a190-321916290c42}. And Chalmers emphasizes that the mathematical methods of reductive physicalism are best suited to describe concepts definable in functional terms\cite{Chalmers1995-CHAFUT}, so it seems difficult to address the quality combination problem using standard mathematical methods. As Seager puts it, `\emph{science deals with the relational structure of the world and consciousness is an, or \textbf{the}, intrinsic feature of the world}'\cite{Seager2010-SEAPAA}. 

On the other hand, relative facts are clearly part of the relational structure of the world, so they can be addressed using the usual methods of physical science: the fact that states and outcomes are relativized to  observers adds some conceptual intrigue, but does not prevent us from defining them with mathematical expressions. Thus the `relative fact combination problem' ultimately comes down to understanding the structural relations between microscopic and macroscopic relative facts, so it can be thought of as  an aspect of the structural combination problem. Therefore I will  focus my discussion of how relative facts combine on the structural combination problem. 

\subsubsection{Structural Combination \label{structural}} 

Evidently the structural combibination problem in RQM  would have a straightforward solution if macroscopic and microscopic relative facts had the same structure, but there is an important physical reason why this probably cannot be the case: the preferred basis problem. That is, in pure unitary quantum mechanics, quantum states  can be represented in many different bases, and there is no physical reason to prefer one over another. And this leads to the following issue. As described in section \ref{intro}, RQM seems to suggest that when an  interaction results in a state which can be written in the Schmidt form in some basis, the interaction creates relative facts in that basis. But as pointed out variously by Pienaar\cite{2021quintet}, Brukner\cite{https://doi.org/10.48550/arxiv.2107.03513} and Muci\~{n}o et al\cite{Mucino2022-MUCARQ}, if the interacting systems are microscopic, there may be multiple bases in which the state takes the Schmidt form. For example, Brukner gives the example of the the entangled state of two qubits $\frac{1}{\sqrt{2}}( | \uparrow \rangle_1 | \uparrow \rangle_1 + | \downarrow \rangle_2 | \downarrow \rangle_2)$, written here using a basis of `spin up' $| \uparrow \rangle$ and `spin down' $| \downarrow \rangle$ states. This state can equivalently be written as $\frac{1}{\sqrt{2}}( | \rightarrow \rangle_1 | \rightarrow \rangle_1 + | \leftarrow \rangle_2 | \leftarrow \rangle_2)$, using a basis of `spin left' $| \leftarrow \rangle$ and `spin right' $| \rightarrow \rangle$ states.  From the first version of the entangled state we infer that there will be a relative fact specifying that qubit two has  either spin up or spin down relative to qubit one, but from the second version of the entangled state we infer that  there will be a relative fact specifying that qubit two has  either spin left or spin right relative to qubit one, so qubit one has relative facts about the value of qubit two's spin in two different bases. 

This seems counterintuitive:  it looks as though qubit one `knows' two mutually incompatible facts about qubit two.  In ref \cite{ pittphilsci19664} di Biagio and Rovelli respond by reinforcing that  it is the interaction rather than the resulting state which determines the basis of the relative facts, so it may be that in cases where there is more than one possible Schmidt form, considering the specific interaction would tell us which one is right. However, it's unclear that this answer will generalize, since as emphasized by ref \cite{mucino2021reply}, interaction Hamiltonians can also be written in different bases, so they will not generically single out one particular basis. And in fact it isn't  necessary to choose a single basis, because the term `observer' is used in  a thin way in RQM, which does not imply the presence of subjective experience.  Thus we can simply say that the qubit in the scenario above does not have knowledge at all, so it doesn't matter if the relative facts associated with it seem incompatible - they are simply  relations which need not be translated into a coherent subjective experience.

But although   the lack of a preferred basis  is not  in and of itself a problem for RQM, it does add complexity to the  structural combination problem.  For in order for RQM to be empirically successful, it must be the case that \emph{macro}-observers  only possess relative facts about a given system in a single basis, since in our actual macroscopic experience we do not ever know incompatible facts about a single system at one time. And this indicates that the structure of the relative facts for macro-observers cannot be inherited directly from the structure of the relative facts for micro-observers.

In fact, there is already a known mechanism for the emergence of a preferred macroscopic basis: decoherence\cite{schlosshauer2007decoherence}. This refers to a process in which quantum systems  rapidly undergo interactions with their environment, with the result that their density matrices  approach a diagonal form in the basis favoured by decoherence, which is typically the coarse-grained position basis. It is often argued that decoherence defines the stable quasi-classical structures corresponding to our macroscopic experience of the world\cite{sep-qm-decoherence}, and so one might naturally think that decoherence should be the reason why macro-observers reliably possess relative facts in just one basis. 

However, it is not automatic that the decoherence mechanism solves the structural combination problem. For example, a solution to the RCP in which we obtain macroscopic relative facts by simply averaging over microscopic relative facts cannot possibly reproduce the effects of decoherence, since  averaging will  result in a space of macroscopic relative facts with the same size and structure as the space of microscopic relative facts. So if we want macroscopic relative facts to actually exhibit the structural features associated with   decoherence, we must build that into our solution to the combination problem, for example by allowing decoherence to play some role in the way in which macroscopic relative facts emerge from the microscopic relative facts. Thus although it certainly seems that decoherence should help with the structural combination problem, we still have to figure out how to properly make use of it.

 \subsubsection{Subject Combination \label{subject}} 

The subject combination problem may be split into two related problems. First, can subjects \emph{ever} combine into an additional subject? And second, if they do combine, under what circumstances do they combine? 

The first problem, pertaining to what Chalmers calls the  subject-summing argument\cite{Chalmers2016-CHATCP-6}, can be articulated in the panpsychist case by imagining a `microexperiential zombie world' in which all physical facts are the same as the actual world, and all fundamental particles have subjective experience, but macroscopic beings do not have subjective experience\cite{Goff2009-GOFWPD}. Goff argues that since this world is conceivable,  the mere existence of  a number of micro-subjects cannot  necessitate the existence of a macro-subject.  Similarly,   we can  imagine an `RQM zombie world' in which the  relative facts associated with  the micro-observers are  the same as in an ordinary RQM world, but there are no macro-observers and no macroscopic relative facts. This seems at least conceivable, for as Mendelovici puts it, `\emph{it is not clearly intelligible why a mere collection of subjects, however organized, should yield a new subject}'\cite{Mendelovici2019-MENPCP}. And if it is the case that no arrangement of micro-observers necessitates the existence of a corresponding macro-observer, then  macro-observers cannot be literally identical to collections  of micro-observers - something further must be added to explain the presence of a macro-observer. This leads to the worry that  we may after all have to treat macro-observers  as an ontologically fundamental category,  putting us at risk of violating the central mandate. 

Now, Goff's response to the micro-experiential zombie argument suggests that `\emph{conceivability entails possibility (only) when you completely understand what you’re conceiving of,}'\cite{Goff2016-GOFTPB} so the conceivability claim in the argument is not philosophically significant unless we fully understand all the relevant concepts, which arguably we do not.  And this line of defence seems quite viable in RQM, since there is surely no reason to think we   fully understand the concept of `observer' in RQM, given that it has been  divorced from its  meaning in ordinary langauge. For example, in a version of RQM+CPL which postulates a substratum of non-relative pointlike events,   it seems likely that certain combinations of micro-observers do necessitate the existence of a macro-observer, since the distribution of events implied by some arrangement of micro-observers could be equivalent to the distribution associated with a certain macro-observer. But we   should not expect the relation between micro-observers and macro-observers to be be revealed  by appeal to a priori conceivability arguments, since   we don't have a grasp of the concepts of micro and macro-observer which extends to knowing which distributions of events they correspond to. 

Nonetheless, the RQM version of the subject-summing argument does at least give us reason to think   that a collection of micro-observers will not \emph{automatically} give rise to a macro-observer; some further component is needed. For example, perhaps the micro-observers only produce macro-observers when they stand in some specific arrangement, as with the patterns of pointlike events. Indeed, Goff reaches a similar conclusion in the panpsychist case, arguing that the zombie argument indicates that `\emph{subjects of experience  cannot sum merely in virtue of their existing}' but there is `\emph{nothing ... which rules out the possibility of there being some state of affairs of certain subjects of experience being related in some specific way which necessitates the existence of some distinct subject of experience}'\cite{Goff2016-GOFTPB}. This suggests that addressing the first part of the combination problem may depend on having a compelling solution to the second. 
 
So what does RQM have to say about the circumstances under which subjects combine? Well, it tells us   that every physical system is an observer, and one might interpret this as saying   that  \emph{every} possible subset of  the fundamental particles in the universe is an `observer,'  with its own set of relative facts.  Some panpsychists have proposed a similar `universalist' solution to their own subject combination problem, suggesting that any  collection of conscious subjects   forms a new conscious subject\cite{Goff2016-GOFTPB}. 

 However, the universalist approach is made less appealing by the subject-summing argument, which suggests that the mere existence of a set of micro-observers is not enough for them to combine into a macro-observer. In addition, universalism in RQM has  odd consequences. First, it implies  that we must associate a set of relative facts with any arbitrary collection of particles, even if they  are far apart and not in causal interaction. And it seems hard to understand the meaning of such relative facts:  these particles cannot act jointly to perform anything that would look like a `measurement,' so  relative facts associated with them cannot be understood as describing the outcomes of any  possible measurements. Second, universalism leads to  a version of the problem of the many\cite{sep-problem-of-many}: there is no way to identify a  unique set of fundamental particles which constitute a human being, so universalism implies that any human being has thousands of different macro-observers associated with her, each defined by a slightly different set of fundamental particles. Furthermore, in the case of orthodox RQM it is likely  that each of these macro-observers will have a  different set of relative facts, and once we allow that, we are starting to get into territory  similar to the Everett interpretation\cite{Everett, Wallace, manyworlds}, since in any measurement performed by an individual human being, all possible outcomes will probably be seen by at least some of her associated macro-observers. Thus insofar as the aim is for RQM to remain clearly distinct from the Everett approach, universalism will be viable only if we can find a way to ensure that all  the macro-observers associated with a given human body share approximately the same relative facts - I will consider one way in which this might be done in  section \ref{combinatorial}.

 A variation on the universalist view would involve identifying the contents of every possible connected region of spacetime as a distinct macro-observer, as Rovelli appears to suggest in ref \cite{Rovelli_2018}. This approach has the notable advantage that it may allow the formulation of RQM without the need for well-defined distinct systems at the fundamental level,  making it more compatible with quantum field theory. However, it retains some of the same problems as universalism based on collections of particles. We will  still end up with many `physical systems' which are not meaningfully unified, and which are not organised  in a way which would allow engagement in joint `measurements' to which the relevant relative facts could apply. And we will still have a version of the problem of the many, since a given sentient being is made up of many different overlapping spacetime regions, which may all have different relative facts. 
 
Given the difficulties with universalism, we might be tempted to define `physical systems' and hence `macro-observers' in a more restrictive way. An alternative which has been suggested in the panpsychist case is nihilism\cite{Goff2016-GOFTPB}, where we say that only  micro-experiences exist, so micro-experiences never constitute macro-experiences.  An RQM version of nihilism would contend that only micro-observers exist and hence micro-observers never constitute a macro-obsever. Evidently this avoids the problem posed by the subject-summing argument - however, it seems challenging to implement in RQM, since the empirical success of the theory relies crucially on attributing relative facts to human observers. With that said, there might be a viable nihilist view in which  it is argued that what we think of as macro-observers are in fact identical to certain micro-observers (I will discuss this in section \ref{identity}), or perhaps one in which we   eliminate macro-observers entirely whilst retaining macroscopic  relative facts (I will discuss this in section \ref{combinatorial}). 
 
Alternatively, we could offer a solution in between universalism and nihilism, in which  some criterion is used to identify specific higher-level systems which count as macro-observers. There are two options here. We  could provide a criterion formulated using macroscopic categories - for example,  perhaps every  sentient being counts as a macro-observer. Or alternatively, we could provide a criterion which can be expressed entirely in microphysical vocabulary - for example, perhaps  every collection of particles which has recently interacted counts as a macro-observer. 

With regard to the first possibility, recall that the identification of macro-observers plays an essential role in identifying the contents of reality according to RQM, so the criterion we use to pick out macro-observers cannot be regarded as merely a linguistic convenience. Therefore if we use a macroscopic criterion which cannot be reduced to microphysical vocabulary, the macroscopic categories in the criterion must be regarded as ontologically fundamental. So  if the criterion uses terms like  `sentient being,' or measurement' we are   in danger of violating the central mandate. 

Moreover, even if we  avoid these particular terms, many other macroscopic notions we might draw on are  subject to semantic vagueness, meaning that  there exist a range of equally acceptable ways of making them  precise\cite{Russell1923-RUSV-2,Dummett1978-DUMTAO}. For example, `spatial proximity' or `a high degree of causal connectivity,' can be made precise in various ways. So if macro-observers are identified using such terms, then the category `macro-observer' itself will be subject to semantic vagueness. Yet relative facts are described by quantum mechanics, which  is a  precise, quantitative physical theory, so there does not appear to be  conceptual space for relative facts  to be  approximately or inconclusively present, thus it is implausible that there could be multiple ways of making precise the nature of the physical system to which relative facts are assigned. Therefore   it seems that the category `macro-observer' should not be subject to semantic vagueness, so even a macroscopic criterion which does not violate the central mandate may still be problematic if it involves semantic vagueness.  

Thus there are good reasons to prefer a criterion expressed entirely in \emph{microphysical} vocabulary, and in sections \ref{emergence} and \ref{combinatorial} I will consider some ways in which this might be done.

\section{Solutions to the Combination Problem \label{solutions}}

Our analysis has yielded the good news  that RCP  is more tractable than PCP, because the problem can be addressed with ordinary mathematical methods. So the range of possible solutions to the RCP is probably larger than the range of possible solutions to the PCP. However, since there is already a significant literature on the PCP, this is nonetheless a good place to start thinking about how to address the RCP. 

Chalmers distinguishes four main classes of solutions to PCP\cite{Chalmers2016-CHATCP-6}: 

\begin{itemize}

\item \textbf{Identity:} macroexperiences are identical to microexperiences.

\item \textbf{Autonomy:}  macroexperiences are autonomous from microexperiences.

\item \textbf{Strong  emergence:} macroexperiences are strongly emergent from microexperiences and are not constituted by them.

\item \textbf{Combinatorial:} microexperiences collectively constitute macroexperiences.

\end{itemize}

\subsection{Identity  \label{identity}}

 One version of the identity solution for panpsychism involves saying that the experiences of a macrosopic subject are identical to the experiences of a single microscopic subject - for example, Leibniz's monadology is a historic example\cite{LeibnizManuscript-LEIM-3}.  A similar approach to RCP might suggest that the set of relative facts for a given macro-observer is identical to the set of relative facts defined relative to some micro-observer. 

However, as noted in section \ref{structural}, the  set of relative facts defined for a micro-observer does not seem to have the same size and structure as a set of relative facts defined for a macro-observer. Thus in order for this approach to solve the structural combination problem,  we would either need to deny what seem to be obvious facts about the structure of our macroscopic experience, or we would need some solution to the preferred basis problem which works at the microscopic level. 

In addition, this approach  does not really solve the subject combination problem in a satisfactory way. Should we say that there is only one macro-observer associated with a given human being, so her relative facts come from just a single particle in her body? This would seem arbitrary, and would have alarming consequences for human mortality - what happens when the particle to which her relative facts belongs departs from her body? But the alternative is to say that there are many macro-observers associated with her, one for each of the particles in her body, and as noted in section \ref{subject} this is likely to collapse into an Everett-style view.

A different version of the identity solution for panpsychism is known as the combinatorial infusion view\cite{Seager2010-SEAPAA}, such that `\emph{when microexperiences combine to yield macroexperiences, they fuse together and cease to exist independently}'\cite{Mendelovici2019-MENPCP}. This could proceed by invoking   quantum holism\cite{Miller2016-MILQH,Ismael2020-ISMQHN-2} in order to say that any systems sharing entanglement should in fact be regarded as a single entity: `\emph{A panpsychist might speculate that such an entangled system, perhaps at the level of the brain or one of its subsystems, has microphenomenal properties. On the quantum holism version of identity panpsychism, macrosubjects such as ourselves are identical to these fundamental holistic entities, and our macrophenomenal properties are identical to its microphenomenal properties'}\cite{Chalmers2016-CHATCP-6}. It is obviously tempting to propose a similar solution to RCP - perhaps  microscopic particles have their own relative facts when they are not  entangled with anything, but once they become entangled they merge into a single entity, so there is just one set of relative facts for a whole entangled macro-observer. 

However, this approach also does not seem to offer the right kind of solution to the subject combination problem. First,  RQM as standardly formulated does not treat entangled systems as a single entity. Indeed, the description of the formation of relative facts given in  section \ref{intro} entails that that there are facts about $Y$ relative to $X$ only when $Y$ and $X$ are entangled from the point of view of an external observer, so if we merge all entangled particles into a single  observer we will not have any relative facts left at all. Second, in a model like RQM without wavefunction collapse,  entanglement is generic: by this point in the history of the universe everything, or at least everything in the vicinity of the Earth, is likely to be highly entangled with everything else. So  it's likely that we would end up with just one big macro-observer in this picture, and relational quantum mechanics with just one observer is surely not meaningfully relational any longer. Thus it seems unlikely that merging observers using entanglement is a viable route to a view which could still be regarded as a version of RQM.

\subsection{Autonomy \label{autonomous}}

An autonomous solution for RCP would   postulate  that macroscopic relative facts are completely autonomous from  microscopic relative facts. However, this kind of approach seems impossible in RQM+CPL, because  CPL ensures that if a macro-observer measures a microscopic particle in an appropriate basis, the relative fact about the outcome of that measurement for the macro-observer will depend on the relevant relative fact for the particle.  Moreover, CPL is likely  to apply to many interactions outside of what we would conventionally consider a `quantum measurement,' so it implies the existence of many connections between macroscopic relative facts and microscopic relative facts. Thus it seems unlikely that there is any  feasible way for the two kinds of facts to be autonomous in RQM+CPL.

On the other hand, one might initially think that in orthodox RQM \emph{only} identity solutions or autonomous solutions are possible,  since the orthodox view does not allow any systematic connections between the relative facts associated with distinct observers. In section \ref{combinatorial} I will argue that there is possibly a way out of this, but nonetheless, clearly the autonomy view is a natural route to take in orthodox RQM. 

However, there is a problem with autonomous solutions. For their answer to the subject combination problem is to simply deny that  macro-observers are related in any way to micro-observers, which means they  necessarily treat macro-observers as fundamental. Therefore they must regard sentient beings as fundamental, since sentient beings are a sub-class of macro-observers. So autonomous solutions    appear to be in danger of violating the central mandate. 

With that said, the central mandate specifically requires that we must avoid regarding sentient beings as a fundamental ontological \emph{category}, and this does not  necessarily mean we cannot regard sentient beings as fundamental per se; they simply cannot be set apart for any special role.  So this  approach could still be consistent with the central mandate if we adopt the universalist approach in which every collection of particles is a macro-observer, or if we can come up with a criterion singling out macro-observers in a way which does not give sentient beings any special role. We have already reviewed the undesirable consequences of the universalist approach, so let us focus on the latter option. 

As discussed in section \ref{subject}, it seems preferable to identify  `macro-observers' using a microscopically-defined criterion, rather than one which refers to macroscopic categories. But here we encounter another obstacle. For if macroscopic relative facts are completely autonomous from microscopic relative facts, then microscopic relative facts are not necessary for the existence of macroscopic relative facts. Yet orthodox RQM tells us that there are   \emph{no} non-relative facts, so if microscopic relative facts are not necessary for the existence of macroscopic relative facts, then the existence of the  microscopic world must be irrelevant to the existence of macroscopic relative facts. This would entail that macro-observers cannot possibly be identified using a microscopically-defined criterion, for if they were so identified, then they and their relative facts \emph{would} depend on the microscopic world. So in this context it seems we have little choice but to identify macro-observers using a macroscopically-defined criterion, which may  make it hard to avoid semantic vagueness and to  fulfil the central mandate.  

Indeed, at this point ontological economy would seem to suggest that we should just deny the existence of the microscopic world altogether, since it is completely unconnected from the macroscopic world and therefore can play no useful role in explaining our macroscopic experience. Moreover, once we do this it looks hard to avoid collapsing into a position which recommends believing only in the existence of sentient observers, or perhaps even into solipsism. Therefore autonomous solutions do not seem to solve the subject combination problem in an adequate way, since they stray too far from the original motivation for RQM in terms of the desire to avoid anthropocentricism.

\subsection{Strong Emergence \label{emergence}}

  The term `strong emergence' is often used to refer to a case in which  emergent entities  cannot be derived directly from the fundamental entities and the fundamental laws alone, but can be derived with   some additional contingent laws of nature  connecting fundamental entities to emergent entities. For examine, Chalmers describes a possible  solution to the PCP in which `\emph{there (are) contingent laws of nature connecting microexperience (or microphysics) to macroexperience'}\cite{Chalmers2016-CHATCP-6}. And we can imagine a similar approach to the RCP, in which there are contingent laws of nature connecting microscopic relative facts to macroscopic relative facts, meaning that the macroscopic relative facts are strongly emergent. 

However, it's unclear that this can be made to work for orthodox RQM. For as Chalmers puts it,  `\emph{Strongly emergent entities and properties are best construed as fundamental entities and properties, not grounded in the base entities or in other entities}'\cite{Chalmers2016-CHATCP-6}. Thus since macro-observers in a strongly emergent approach are not grounded in micro-observers, they must  be treated as distinct observers. But this means that according to orthodox RQM they cannot share any information with micro-observers, which means they must in fact be completely autonomous from the micro-observers, rather than strongly emergent. Thus for orthodox RQM this approach would seem to collapse into the autonomy approach. 

On the other hand, strong emergence seems feasible for RQM+CPL, since it permits systematic connections between relative facts for different observers. We would need to define the laws of nature connecting microscopic and macroscopic relative facts in the right way to solve the structural combination problem, but that seems achievable, particularly if we make use of decoherence. However, since strongly emergent macro-observers are regarded as fundamental, before we can postulate laws connecting microscopic relative facts to macroscopic relative facts we will need a solution to the subject combination problem, i.e. we will need a criterion which tells us how to identify the macro-observers in the first place. 

Evidently the nihilist approach will not work for a strong emergence picture, and meanwhile we have seen that the universalist approach has various disadvantages, so we will probably want to take the intermediate approach by giving a criterion which identifies certain specific complexes of particles as macro-observers. Moreover, as argued in section \ref{subject},  it's preferable for this criterion to be microscopically-defined. What might a microscopically-defined criterion look like?
 
We could perhaps appeal to entanglement, so for example we might stipulate that a `macro-observer' is a complex of particles sharing entanglement. However, as noted earlier, at this point in history it is likely that entanglement is very widespread, so simply identifying \emph{any} collection of entangled particles as a `macro-observer' would probably be casting the net too wide. To avoid this we could  insist that the complex shares a sufficiently high level of mutual entanglement, but then the criterion would either be subject to semantic vagueness, which  does not seem acceptable in the definition of a macro-observer, or it would have to be precisified in an unsatisfactorily arbitrary way. Furthermore, entanglement by itself cannot transfer information, so particles which are entangled but widely separated cannot share information or coordinate action, so it's not clear that they could perform any joint `measurement' to which the relative facts associated with them could apply. 

Alternatively, we could use a criterion based on information-sharing  using the CPL mechanism, which would ensure that `macro-observers' are meaningfully unified and potentially capable of shared action.  However, at that point it's no longer clear that we need \emph{strong} emergence, since a unified perspective would seem to emerge through this mechanism even without the addition of special laws of nature connecting microscopic relative facts to macroscopic relative facts. So this approach would potentially work better as a combinatorial solution, as I will discuss in section \ref{combinatorial}.

In addition, the strong emergence solution for RCP faces a similar problem to the strong emergence solution for PCP: `\emph{the motivation leading to panpsychism in the first place appears to evaporate. If ontological emergence is possible, then why not simply accept that consciousness emerges in this — radical— way from a purely physical, entirely nonexperiential, fundamental substrate}'?\cite{Seager2010-SEAPAA}. Likewise, if macroscopic observers and relative facts can be derived by strong emergence in a way that does not violate the central mandate, what is the motivation for postulating microscopic relative facts at all? It would  be more ontologically economical to just let the macroscopic observers and relative facts emerge from some non-relational substratum, thus eliminating a large network of microscopic relative facts. So it's likely that we will be better off taking the combinatorial approach if we wish the result to remain a version of RQM.

\subsection{Combinatorial Solutions \label{combinatorial}}

A combinatorial solution for RCP would  postulate  that microscopic relative facts collectively constitute macroscopic relative facts, or that macroscopic relative facts \emph{weakly} emerge from microscopic relative facts, possibly together with other microphysical facts.  One might initially think that combinatorial solutions are impossible in orthodox RQM, since it permits no systematic connections between the relative facts of distinct observers. However, note that if we adopt a  view in which  macro-observers are literally  constituted out of micro-observers,  they are not strictly speaking   distinct  any longer. So  there may be a reasonable version of orthodox RQM in which we allow  systematic connections between the relative facts of a macro-observer and the relative facts of the micro-observers that constitute it, even though there cannot be any systematic connections between the relative facts of distinct observers. 

Still, in such an approach we would  be quite limited in the way in which we  derive macroscopic relative facts from microscopic relative facts. For in the orthodox picture we cannot allow distinct micro-observers to share any information, and therefore it is impossible to make use of any \emph{dynamical} or diachronic process, like decoherence, to create relations  between the relative facts of different micro-observers which would explain the presence of  a macro-observer: the relative facts associated with individual micro-observers will always be completely independent. Thus instead of appealing to some kind of dynamical energence, we will have to adopt a synchronic approach in which the values of the relative facts for a macro-observer at a given time are obtained directly from some function of the relative facts for each of the independent micro-observers constituting it at that time -  for example,   some kind of averaging or `voting'  process.

Such an approach would potentially  solve some  certain aspects of the subject combination problem. For example, we might still have to say that there are many different overlapping observers associated with each human body, but if the relative facts for these macro-observers are obtained from the micro-observers by averaging or voting, significantly overlapping observers will usually have the same relative facts, so we will no longer be in an Everett-style situation with different versions of the same individual seeing different outcomes for a given measurement. 

However, we may still have a problem with the subject-summing argument. For we have noted that it seems implausible that the mere existence of a set of micro-observers  will on  its own produce a macro-observer. Something more is needed - for example, the micro-observers may need to stand in some specific kind of relation to each other. But in the orthodox approach, although it may be the case that for any given observer $O$ other observers stand in some relation to $O$ and to each other relative to the perspective of $O$, those are only facts relative to $O$ - the relations may look completely different relative to other observers, and in general there will not be any structure shared between all the perspectives which could ground the existence of a macro-observer. Indeed, the idea that the presence of a macro-observer requires something over and above the existence of the individual micro-observers seems to suggest that there must be some \emph{non-relative} fact about the relations between the observers which binds them together into a macro-observer. But this is not not allowed in orthodox RQM, so the subject-summing issue remains problematic. 

There are also difficulties with the structural combination problem, because we know  the structure of the macroscopic relative facts should  not be the same as the structure of the microscopic relative facts, whereas simple functions like averaging and voting tend to keep the overall structure the same. The obvious way to arrive at the right macro-structure is by appeal to decoherence, but since the orthodox picture can't make use of dynamical processes in the production of macro-observers, we would just have to put the effects of decoherence in by hand, which seems inelegant. Moreover, as soon as we employ a function more complex than  averaging or voting, this approach starts to look more like strong emergence than constitution or weak emergence, with the chosen function playing the role of a fundamental law connecting microscopic relative facts and macroscopic relative facts - and  in a strong emergence approach macro-observers must be regarded as distinct from  micro-observers, in which case orthodox RQM entails that there cannot after all be systematic links between their respective relative facts! 

On the other hand, the situation seems much better in RQM+CPL, since it lets us make use of dynamical processes to create a unified perspective via the sharing of relative facts.   However,  this approach might still face a problem related to the  subject summing argument: even if a collection of  micro-observers comes to share some relative facts, that doesn't by default produce any higher-level macro-observers or any higher-level relative facts. For example, suppose a rumour goes around a crowd until everyone knows the rumour; now there is shared knowledge, but we would not normally say that there is some new collective entity to whom that knowledge belongs, over and above all the individual people. And similarly in the case of RQM+CPL, it seems something more must be added if we want the sharing of information to give rise to a macro-observer with its own relative facts, over and above all the individual micro-observers. 
 
Here we may take inspiration from a  a popular version of the combinatorial solution in panpsychism. As Chalmers puts it, `\emph{the subject-summing argument is generated in part by thinking of microsubjects as being merely related spatiotemporally or causally. Once we acknowledge distinctively phenomenal relations between microsubjects and their phenomenal states, we can see how all this might constitute a macrosubject and macrophenomenal states}'\cite{Chalmers2016-CHATCP-6}. For example, Goff proposes a phenomenal bonding relation `\emph{which bonds subjects together to produce other subjects of experience}'\cite{Goff2016-GOFTPB}. Various proposals exist for the specific nature of this relation - for example  Chalmers imagines a `\emph{co-consciousness relation ... (such that) when this relation holds among the states of distinct microsubjects, those states will be experienced jointly by a new subject}'\cite{Chalmers2016-CHATCP-6}.

We can imagine something similar in RQM+CPL. For example, suppose we define an `informational bonding relation'  which holds between any any pair of observers which have come to share a relative fact via an interaction as set out in the CPL postulate. Then we can stipulate that whenever two observers are related by an informational bonding relation involving the fact $F$, there is also a composite  observer present, in possession of the same relative fact $F$. This  circumvents the  subject-summing argument, since it adds the informational bonding relation to bring macro-observers and higher-level relative facts into being, and it does this in a way which is compatible with the central mandate, since  the informational bonding relation  is defined using microphysically precise vocabulary.

 Note that when a pair of observers $X$, $Y$ form an informational bonding relation involving the relative fact(s) $F$, we should attribute to the resulting higher-level observer $Z$ only the relative fact(s) $F$ which $X$ and $Y$ share. We cannot attribute to $Z$ \emph{all} of the relative  facts associated with both $X$ and $Y$, since it is possible that some of the facts which have not  been shared are contradictory or incompatible. This suggests that higher-level observers will often have fewer relative facts than the observers they are constituted from.

And in fact, combining this information-losing feature with decoherence gives exactly the result we need to solve the structural combination problem. For an informational bonding relation to form, $X$ and $Y$ must interact such that their entangled state can be written in the Schmidt form $\sum_i  | U_i \rangle_X \otimes | V_i \rangle_Y$, and then $X$ and $W$ must interact such that their entangled state can be written in the Schmidt form $\sum_i  | U_i \rangle_X \otimes | Q_i \rangle_W$, giving rise to an informational bonding relation which produces an observer composed from $X$ and $W$, in possession of a relative fact about $Y$ in the basis $V$.  Moreover, decoherence ensures that as the systems involved become large, effectively every interaction will produce a state which can be written in the Schmidt form $\sum_i  | B_i \rangle_X \otimes | B_i \rangle_Y$, where $B$ is the basis favoured by decoherence\footnote{In fact decoherence is dynamical and the state will only approach this form asymptotically, but RQM is concerned with the idealization of a `complete' observation, so we can assume the exact form $\sum_i  | B_i \rangle_X \otimes | B_i \rangle_Y$.}. Thus whenever any larger system interacts with two different systems, this will effectively \emph{always} result in an informational bonding relation involving a relative fact defined in the basis $B$, so it will  produce a composite observer possessing a relative fact defined in the basis $B$. Whereas in cases where a relative fact is  also  produced  in another basis, it is unlikely that the next interaction will also involve an additional relative fact in a matching basis, so composite observers with relative facts in bases other than $B$ are unlikely. Therefore by the time we arrive at a sentient being constructed from millions of informational bonding relations, it's fantastically unlikely that it will possess relative facts in any basis other than the one favoured by decoherence. So this approach has the welcome result that  the decoherence mechanism plays an active role in giving us exactly the macroscopic structure we need to explain our macroscopic experience. 

However, there is a difficulty for the phenomenal bonding approach   which may apply here as well. As Chalmers puts it: `\emph{One question for this view and for other phenomenal bonding views is whether the bonding relation is transitive  ... If so, then given the ubiquity of spatiotemporal and causal relations, it looks as if the microphenomenal states throughout the universe may stand in this relation, yielding a single giant subject}'\cite{Chalmers2016-CHATCP-6}. And a similar problem may arise in RQM. The informational bonding relation itself is not transitive, since if $X$ shares $F$ with $Y$, and $Y$ shares $F'$ with $Q$, it does not follow that there is any relative fact they all share. However,  the relation of `sharing some \emph{specific} piece of information' is transitive, which looks problematic   - for while there may not be   any piece of information shared by literally the entirety of the human race, there are certainly pieces of information which are shared by very large sections of the human race, so it seems that in this approach most of the human race will be unified into one giant `observer.' 

  However, recall that  `observer' in RQM does not  mean `subject of conscious experience.' We have noted already that micro-observers such as fundamental particles need not have conscious experiences, but on the flip side, there is also no requirement that a `macro-observer' is associated with just one conscious experience. So  there may be nothing wrong with saying that a whole community is a single `observer,' since this just means that there are some relative facts shared by the entire community.  Indeed, as argued in refs\cite{ https://doi.org/10.48550/arxiv.2203.13342,https://doi.org/10.48550/arxiv.2203.16278}, one of the main reasons for moving from orthodox RQM to RQM+CPL  is to affirm the possibility of  sharing  information in the process of doing science, so from this point of view the  merging of the whole epistemic community into a single `observer' is actually a positive.

Of course, in a realistic community there will be some relative facts shared by everyone, and then also many relative facts possessed by various subsets of the community; so in addition to one big observer we can expect to obtain many smaller overlapping `observers,' each covering different segments of the community. One conclusion we might draw from this is that the term `observer' may not be so useful in RQM+CPL. For once we  detach the term from  conscious experience, and we allow exchange of information between observers, the category `observer' is not doing much other than tracking distributions of relative facts across various collections of particles, and  it's those distributions rather than  facts about `observers' which represent the substantive content of the theory.

So perhaps what we ultimately end up with in this approach is an eliminativist view -  we are still employing a combinatorial approach to derive macroscopic relative facts from microscopic relative facts, but we are  eliminating macro-\emph{observers},  thus rather than solving the subject combination problem, we are  defusing it. In particular, this approach straightforwardly sidesteps difficulties in the style of the `problem of the many,' because it entails that counting macro-observers associated with a human body is not a meaningful thing to do:  we should just focus on the  relative facts, and we will typically find that all the subsystems of the human body share approximately the same relative facts, thus accounting for the unified perspective of a human agent. 

Indeed, Chalmers notes that eliminativism about higher-level subjects is a straightforward solution to the panpsychist subject combination problem\cite{Chalmers2016-CHATCP-6}, but is usually dismissed because most people struggle to believe that there are no macroscopic subjects of conscious experience. By contrast, the technical notion of `observer' in RQM is really just a locus to which relative facts are attached, so most people presumably would not have a strong intuition that  there \emph{must} exist observers in this sense, over and above bundles of relative facts.  Indeed, Oldofredi proposes a somewhat similar view  in ref \cite{Oldofredi_2021}, though since this is couched within orthodox RQM rather than RQM+CPL it does not address the combination problem. Thus in RQM+CPL the motivation for the eliminativist view looks comparatively much stronger.

\section{Conclusion}

We have seen that orthodox RQM runs into a central difficulty with the combination problem. For if macroscopic relative facts are autonomous from microscopic relative facts, it seems quite difficult to to satisfy the central mandate. But the orthodox approach  cannot allow macroscopic relative facts to be related to microscopic relative facts unless we insist that the macro-observers are not distinct from micro-observers; and in this case the macroscopic relative facts will likely have to be determined as a relatively simple function of a set of completely independent microscopic relative facts, in order to avoid collapsing into a strong emergence picture in which the macro-observers are not after all distinct from the micro-observers. Furthermore, it's unclear that any sufficiently simple function will be capable of solving the structural combination  problem in a way that respects our physical understanding of the differences between microscopic and macroscopic structure. So we seem to be pushed towards the conclusion that in orthodox RQM, macroscopic relative facts must indeed be autonomous from microscopic relative facts, and therefore we may not be able to fully articulate orthodox RQM without treating sentient beings or some similar macroscopic category as  ontologically fundamental.

The prospects for solving the combination problem in RQM+CPL seem significantly better, because the CPL mechanism allows us to have connections between the relative facts of distinct observers, so we can  use  dynamical processes like decoherence to inform the emergence of macroscopic relative facts from microscopic relative facts. The `informational bonding' approach  seems like a promising way to combine  microscopic relative facts using only mathematically precise, microscopically-definable notions, and the idea that having a unified perspective is connected to the ability to share information seems fairly natural. Thus, since   solving RCP seems essential  to RQM's overall goals, the improved prospects for a solution to RCP  offers a further reason to prefer RQM+CPL over the orthodox approach.

\end{document}